\documentclass[12pt]{iopart}
\usepackage{graphicx}

\usepackage{hyperref}
\usepackage{fancyhdr}
\begin{document}

\title[Nanocrystal Energetics via Quantum Similarity Measures]{Nanocrystal Energetics via Quantum Similarity Measures}

\author{M. \.{I}. T\"{o}rehan Balta and \c{C}etin K{\i}l{\i}\c{c}\footnote{Author to whom any correspondence should be addressed}}
\address{Department of Physics, Gebze Institute of Technology, Gebze, Kocaeli 41400, Turkey}
\ead{cetin\_kilic@gyte.edu.tr}

\begin{abstract}
We first develop a descriptor-based representation of atomic environments
  by devising {\it two\it} local similarity indices
  defined from an atom-partitioned quantum-chemical descriptor.
Then we employ this representation
  to explore the size-, shape-, and composition-dependent nanocrystal energetics.
For this purpose we utilize an {\it energy difference\it} $\mu$
  that is related to the atomic chemical potential,
  which enables one to characterize energetic heterogeneities.
Employing first-principles calculations based on the density functional theory
  for a set of {\it database\it} systems, viz.
  unary atomic clusters in the shape of regular polyhedra and the bulk solids of C, Si, Pd, and Pt,
  we explore the correlations between the energy difference $\mu$ and similarity indices.
We find that there {\it exists\it} an interconnection between
  nanocrystal energetics and quantum similarity measures.
Accordingly we develop a means for computing total energy differences
  from the similarity indices via interpolation,
  and utilize a {\it test set\it} comprising a variety of unary nanocrystals and binary nanoalloys/nanocompounds
  for validation.
Our findings indicate that the similarity-based energies
  could be utilized
  in computer-aided design of nanoparticles.
\end{abstract}


\maketitle

\section{Introduction}
Nanoparticles exhibit chemical and structural inhomogeneities
  insomuch as they appear in various morphologies,
  regardless of the production method used\cite{Trindade2001,Ferrando2008,Xia2009}.
Even a well-crystallized nanoparticle with a well-defined polyhedral shape
  consists of low-coordinated surface atoms
  that are obviously dissimilar to high-coordinated (bulk-like) atoms.
One recognizes immediately
  by inspection
  that the bulk-like atoms have indeed a local (bonding) environment that resembles the {\it bulk solid\it}
  whereas the surface atoms along the edges or at the corners have similarities with the atoms of {\it atomic clusters\it}
  in regard to bonding and local coordination.
Utilizing these similarities (or dissimilarities) 
  in elucidating the properties of nanoparticles
  would be rewarding since one could then employ similarity search methods\cite{Willett1998,Sheridan2002,Engel2006}
  in computer-aided design of nanoparticles with customized properties.
We should like to complement this consideration
  by noting that
  similarity measures/indices\cite{Carbo1980,Mezey1999,Chermette1999}
  based on quantum-chemical descriptors\cite{Karelson1996,Bultinck2005,Geerlings2005}
  have long been available,
  which proved to be promising
  in carrying out various tasks
  related to a multitude of physicochemical phenomena,
  such as comparing properties and reactivities of different molecular systems\cite{Boon2001},
  deriving quantitative structure-activity or -property relationships\cite{Good1993,Besalu2002},
  and identification of the active molecular sites\cite{Amat2001}.
Clearly these efforts demonstrate the utility of 
  the similarity-based analysis in molecular design\cite{Rouvray1992}.
On the other hand,
  the structure and/or properties of nanoparticles
  have {\it never\it} been explored via the notion of the (quantum) similarity.
This is, in our opinion, due to lacking a {\it complete\it}\cite{Bartok2013} similarity-based representation of atomic environments,
  which furnishes an adequate description for the nanoparticle atoms.
The present study is thus devoted to fulfill an objective along this line:
First, we develop
  a descriptor-based representation of atomic environments
  by devising {\it two\it} local similarity indices
  that are defined from the atom-partitioned\cite{Hirshfeld1977,DeProft2003} {\it shape function\it}\cite{Parr1983,Ayers2000,Proft2004} $\sigma(\mathbf r)=\rho(\mathbf r)/N_e$,
  where $\rho(\mathbf r)$ and $N_e$ denote the electronic density function and number of electrons, respectively.
Then, we employ this representation
  to explore the size-, shape-, and composition-dependent nanocrystal energetics
  via the notion of quantum similarity.
We focus on an {\it energy difference\it} $\mu$,
  which is related to the {\it atomic chemical potential\it},
  for its utility in the modelling and simulation of nanoparticles\cite{Kilic2010,Kilic2011}.
This energy difference and the local similarity indices
  were obtained by performing first-principles calculations based on the density functional theory (DFT) for
  (i) a set of {\it database\it} systems including unary atomic clusters in the shape of regular polyhedra and the bulk solids of C, Si, Pd, and Pt,
  and
  (ii) a {\it test set\it} for validation, which includes a variety of unary nanocrystals as well as binary Pt-Pt nanoalloys and Pt-C nanocompounds.
Regarding the energy difference $\mu$ as a ``property''
  and exploring its correlations with the local similarity indices,
  we find that there {\it exists\it} an interconnection between
  nanocrystal energetics and quantum similarity measures.
Furthermore, we introduce an interpolation procedure in order to obtain total energies and energy differences
  from the two similarity indices.
The latter enables us to demonstrate that the similarity-based energies
  could be utilized
  in computer-aided design of nanoparticles.

Although one of the similarity indices (denoted by $Z_{i\alpha}$ below) employed in this study is of the same form as
  the local Carb\'{o} index\cite{Carbo1980,Mezey1999,Geerlings2005},
  we find it necessary to introduce a {\it second\it} similarity index (denoted by $S_{i\alpha}$ below)
  in order to achieve an adequate chemical representation.
Our findings reveal that using a single similarity index, cf. $Z_{i\alpha}$,
  leads to a bijection between $Z_{i\alpha}$ and $\mu$
  {\it if and only if\it} the set of systems are restricted to include {\it equilibrium\it} structures.
This has the obvious drawback that
  it requires {\it a priori\it} knowledge of the {\it equilibrium\it} geometries.
Thus we find the introduction of a second similarity index necessary,
  which enables us to extend the aforementioned bijection to cover a generalized set of systems,
  including strained (compressed or dilated) systems.
Hence the foregoing drawback is overcome by using {\it two\it} similarity indices
  with adequately tailored functional forms given below.
It should be emphasized that this approach makes a similarity-based representation of the potential energy surfaces\cite{Behler2011} accessible
  thanks to the inclusion of the strained systems,
  which facilitates an adequate description of physicochemical processes.

The rest of the paper is organized as follows:
The next section is devoted to the methodological aspects
  which also summarizes the computational details.
This is followed by a discussion of the calculation results
  before concluding remarks given in the last section.

\section{Methodology}

In this section we first define the energy difference $\mu$ and the similarity indices that constitute the employed chemical representation.
We then describe (i) the set of {\it database\it} systems that are used to construct a database
  for the purpose of exploring correlations between energy differences and local similarity indices, and 
  (ii) the {\it test set\it} employed for validation.
Next we introduce an {\it ad hoc\it} interpolation procedure that enables one to compute the similarity-based energy differences and total energies.
We finalize this section with a description of our computational modelling framework.

\subsection{Energy differences}
The aforementioned energy difference is defined by
  \begin{eqnarray}
    \mu(N,d) = \Delta H_{\rm a} + \left [ E_N(d) - E_{\rm b}(d_b) \right ],
  \label{eqMu}
  \end{eqnarray}
  where $N$ is the number of atoms in the system under consideration,
  $E_N(d)$ is the DFT-calculated energy per atom for the system with the nearest-neighbor distance $d$,
  $E_{\rm b}(d_{\rm b})$ is the DFT-calculated energy per atom for the bulk solid with the {\it equilibrium\it} nearest-neighbor distance $d_{\rm b}$
  corresponding to the minimum of the total energy,
  and $\Delta H_{\rm a}$ denotes the negative of the measured\cite{CRC} {\it heat of atomization\it} for the bulk solid.
It is useful to define $\mu_{\rm e} (N) = \mu (N,d_e)$
  for the {\it equilibrium\it} value $d_e$ of the nearest-neighbor distance.
Note that
  $\mu_{\rm e} (N)$ would be equal to the {\it atomic\it} chemical potential at zero temperature ($T=0~{\rm K}$)
  for a {\it unary\it} system
  if the {\it zero of energy\it} is set to the energy of the atom,
  i.e., $\mu_{\rm atom}=0$.
Recent investigations\cite{Kilic2010,Kilic2011} by one of the present authors
  have indicated that the energy difference $\mu$ could be utilized to introduce a {\it scale\it} of energy
  on which small (less stable, more reactive) and large (more stable, less reactive) nanocrystals
  are naturally ordered near the higher ($\mu_{\rm atom}$) and lower ($\mu_{\rm bulk}$) ends of the scale, respectively.

\subsection{Local similarity indices}
In the present study,
  the isolated (free) atom is employed as a {\it reference\it} system for each atomic species:
  namely, the free C, Si, Pd, or Pt atoms are used as a reference for the C, Si, Pd, or Pt atoms in any system,
  viz. atomic clusters, bulk solids, or nanocrystals, respectively.
This is advantageous
  for computational efficiency and avoids the need for an alignment\cite{Parretti1997,Bultinck2003,Bultinck03} procedure.
It implies that the similarity of an atom of a certain type in a system is measured
  with respect to the free atom of the same type,
  regardless of the type of the system (atomic cluster, bulk solid, or nanocrystal).
The local (atom-partitioned) Carb\'{o} index,
 which here serves as an indicator of similarity of an atom $i$, located at $\mathbf R_i$, of type $\alpha$ in the system under consideration to the free $\alpha$ atom,
 could then be expressed as
  \begin{equation}
    Z_{i\alpha}=\frac{\int \sigma(\mathbf r) w_i(\mathbf r-\mathbf R_i) \sigma_\alpha(\mathbf r)d\mathbf r}{\sqrt{\int  \sigma^2(\mathbf r)w_i(\mathbf r-\mathbf R_i)d\mathbf r}\sqrt{\int \sigma^2_\alpha(\mathbf r)d\mathbf r}},
  \label{eq1st}
  \end{equation}
  where $\sigma(\mathbf r)$ and $\sigma_\alpha(\mathbf r)$ denote
  the shape function of the system under consideration and free $\alpha$ atom, respectively.
The use of the Hirsfeld partitioning\cite{Hirshfeld1977,DeProft2003} function
  $w_\alpha(\mathbf r-\mathbf R_\alpha)=\rho_\alpha(\mathbf r-\mathbf R_\alpha) / \rho_{m}(\mathbf r)$
  in equation~(\ref{eq1st})
  is encouraged\cite{Geerlings2005} by the holographic electron density theorem\cite{Mezey1999}.
Here $\rho_\alpha(\mathbf r-\mathbf R_\alpha)$ denotes the electron density of the isolated atom $\alpha$ located at point $\mathbf R_\alpha$,
  and $\rho_{m}(\mathbf r)=\sum_\alpha \rho_\alpha(\mathbf r-\mathbf R_\alpha)$ is the promolecular electron density.
As explained above, using $Z_{i\alpha}$ does not suffice for obtaining a full-fledged representation of atomic environments.
Thus we introduce a {\it second\it} indicator of local similarity given by
  \begin{equation}
    S_{i\alpha}=\frac{\int r^2 \sigma(\mathbf r) w_i(\mathbf r-\mathbf R_i) \sigma_\alpha(\mathbf r)d\mathbf r}{\sqrt{\int r^2 \sigma^2(\mathbf r)w_i(\mathbf r-\mathbf R_i)d\mathbf r}\sqrt{\int r^2 \sigma^2_\alpha(\mathbf r)d\mathbf r}}.
  \label{eq2nd}
  \end{equation}
As explained below, the introduction of $S_{i\alpha}$ enables one to treat
  {\it energetic trends\it}
  when the variations with the interatomic distance $d$ are taken into account.
Note that $S_{i\alpha}=Z_{i\alpha}=1$ if the system itself is the free $\alpha$ atom.

\subsection{The set of database systems}
In this study, first-principles calculations are employed for building 
  a database that comprises $\mu$ and ($Z_{i\alpha}$,$S_{i\alpha}$) values
  for the {\it unary\it} atomic clusters in the shape of Platonic or Archimedean solids.
In addition to these regular polyhedra,
  dimers C$_2$, Si$_2$, Pd$_2$, and Pt$_2$, and
  the bulk solids of C, Si, Pd, and Pt are included in this database.
It should be emphasized that not only equilibrium systems ($d=d_e$)
  but also {\it compressed\it} or {\it dilated\it} systems ($d < d_e$ or $d > d_e$)
  are included in this set.
The systems in this set are called here {\it database\it} systems for ease of speech,
  which are thoroughly used for the purpose of exploring the correlations between the $\mu$ and $Z_{i\alpha}$ or $S_{i\alpha}$ values.
That 
the set of
{\it database\it} systems comprise only {\it equivalent\it} atoms
  makes it possible to set $E_N=E_{\rm DFT}/N$ in equation~(\ref{eqMu}),
  where $E_{\rm DFT}$ denotes the DFT-calculated total energy.
For each {\it database\it} system,
  plotting $\mu$ values as a function of $Z_{i\alpha}$
  yields a convex curve that could accurately be parameterized, as demonstrated below.
Accordingly, 
 for a given {\it database\it} system $i$,
 the energy difference defined in equation~(\ref{eqMu}) is represented by $\mu=\mu_i(Z_{i\alpha})$
 where $\mu_i$ denotes a polynomial function of forth order
 (whose coefficients are determined by fitting to the DFT-calculated values).
It is found that one must employ a distinct function $\mu_i$ with a unique set
  of polynomial coefficients for each system.
Thus the $\mu=\mu_i(Z_{i\alpha})$ relationships are tabulated for all {\it database\it} systems
  with $\alpha=$ C, Si, Pd, Pt in tables~1-4 in Supplementary Data.

\subsection{Test set}
As mentioned above,
  a variety of nanocrystals
  are utilized as {\it test\it} systems for the purpose of validation,
  which exhibit structural inhomogeneity
  owing to the presence of a number of {\it inequivalent\it} atoms.
This {\it test\it} set is designed
  to cover a variety of nanocrystal sizes and shapes
  as well as a range of nanoalloy compositions with various mixing patterns.
Thus
 a number of unary C, Si, Pd, or Pt nanocrystals (Supplementary Data, table~5),
     uniformly mixed (Supplementary Data, table~6),
     core-shell segregated (Supplementary Data, table~7), and
     phase separated (Supplementary Data, table~8) Pt-Pd nanoalloys, and
     Pt-C nanocompounds (Supplementary Data, table~9)
 are contained in the {\it test\it} set.
In practice, the atoms of the {\it test\it} systems
  were constrained to occupy
  the diamond (C and Si nanocrystals),
      fcc (Pd and Pt nanocrystals and Pt-Pd nanoalloys), and
      zinc-blende (Pt-C nanocompounds) lattice sites.
It should be mentioned that platinum carbide nanocrystals are considered here only for the purpose of
  studying some challenging systems
  since the bonding characteristics of platinum carbide,
  which was synthesized\cite{Ono2005} for the first time in 2005 under extreme conditions
  via a high-pressure and high-temperature method in a diamond anvil cell with laser heating,
 is peculiar owing to the mixed covalent-ionic-metallic\cite{Ivanovskii2009} interatomic interactions.

\subsection{Interpolation procedure}
With the aid of tabulated $\mu=\mu_i(Z_{i\alpha})$ relationships (Supplementary Data, tables~1-4),
 the energies of the {\it test\it} systems are obtained according to the following procedure:
The contribution $\delta E_I$ to the energy by the $I$ atom of a {\it test\it} system
  is obtained, via interpolation, by
  \begin{eqnarray}
    \label{eqEloc}
    \delta E_I &=& \sum_i \omega_{Ii} ~ \mu_i(Z_{i\alpha}),\\
    \label{wii}
    \omega_{Ii} &=& \frac{(S_{I\alpha}-S_{i\alpha})^{-2}}{\sum_i(S_{I\alpha}-S_{i\alpha})^{-2}},
  \end{eqnarray}
  where $\omega_{Ii}$ denotes the interpolation coefficients, and
  $i$ is the label for {\it database\it} systems
  (whereas $I$ denotes atoms of the {\it test\it} system under consideration).
Performing a sum over $I$ yields
  a similarity-based {\it total\it} energy difference
  \begin{equation}
    \Delta E_{\rm sim} = \sum_I \delta E_I.
    \label{eqEsim}
  \end{equation}
Note that
 $\Delta E_{\rm sim}$ should be compared to
 a DFT-calculated {\it total\it} energy difference given by
  \begin{eqnarray}
    \Delta E_{\rm DFT}=m \Delta H_a^{\rm A} + n \Delta H_a^{\rm B}+
                  \left [
                  E_{\rm DFT}({\rm A}_m{\rm B}_n) - m E_b^{\rm A} - n E_b^{\rm B}
                  \right ],
    \label{eqEdft}
  \end{eqnarray}
  for a nanoalloy/nanocompound made of $m$ A and $n$ B atoms,
  owing to the inclusion of $\Delta H_a$ and $E_b$ in the definition of $\mu$, cf. equation~(\ref{eqMu}).
One obviously needs to set $n=0$ in equation~(\ref{eqEdft}) for a unary nanocrystal made of $m$ A atoms.

\subsection{Computational details}
The DFT-calculated energies employed in equations~(\ref{eqMu}) and (\ref{eqEdft})
  as well as shape functions employed in equations~(\ref{eq1st}) and (\ref{eq2nd})
  were obtained
  within the generalized gradient approximation (GGA)
  using the PBE exchange correlation potential\cite{PBE},
  and employing the projector augmented-wave (PAW) method\cite{PAW}, as implemented in VASP code\cite{VASP,VASPpaw,VASPchem}.
Spin-polarization was taken into account and {\it scalar\it} relativistic effects were included
 in all calculations.
The 2$s$ and 2$p$, 3$s$ and 3$p$, 4$d$ and 5$s$, and 5$d$ and 6$s$
states are treated as valence states for carbon, silicon, palladium, and platinum, respectively.
Plane wave basis sets were used to represent the electronic states,
  which were determined by imposing a kinetic energy cutoff of 400, 245, 250, and 230 eV
  for C, Si, Pd, and Pt, respectively.
Primitive and/or conventional unit cells were used in the calculations for the bulk solids, viz.
  C and Si in the diamond structure and Pd and Pt in the face-centered-cubic (fcc) structure,
  whose Brillouin zones were sampled by fine {\bf k}-point meshes
  generated according to Monkhorst-Pack scheme\cite{MP1976},
  ensuring convergence with respect to the number of {\bf k}-points.
A variety of cubic supercells with a side length in the range 15--30 \AA~ were used for
  the atomic clusters and nanocrystals,
  which included a vacuum region that put at least 10 \AA~ distance
  between nearest atoms of two systems in neighboring supercells.
Only $\Gamma$-point was used for Brillouin zone sampling in the case of the cluster or nanocrystal supercells.
The error bar for the energy convergence was on the order of 1 meV/atom in all calculations.

The overlap integrals employed in equations~(\ref{eq1st}) and (\ref{eq2nd})
  were evaluated via an adaptive multidimensional integration routine\cite{dcuhre}
  within a spherical region about the atomic centers in real space.
For efficiency in computing the integrands in equations~(\ref{eq1st}) and (\ref{eq2nd}),
  spline interpolations\cite{spline} of the electron density functions $\rho(\mathbf r)$ and $\rho_\alpha(\mathbf r)$ were performed
  using the three-dimensional gridded data written by VASP.
The integration region for any atom $\alpha$ was imposed
  by setting the integrands to zero at every point $\mathbf r$ where $\rho_\alpha(\mathbf r) < 3\times10^{-6}$ e/\AA$^3$.
That this approach yields sufficiently accurate results
  was checked by computing the normalization integrals such as $\int \sigma(\mathbf r) d\mathbf r = 1$
  and also the integrals such as $\int \sigma(\mathbf r) w_i(\mathbf r-\mathbf R_i) d\mathbf r$ which should yield $1/N$
  for the atomic clusters in the shape of Platonic or Archimedean solids.
For infinite systems such as bulk solids
  (for which the shape function is zero everywhere but normalized to unity\cite{Ayers2000})
  one could still apply this approach
  thanks to
  the inclusion of the denominator terms in equations~(\ref{eq1st}) and (\ref{eq2nd}), and
  {\it spatial localization\it} imposed by the partitioning function $w_i(\mathbf r-\mathbf R_i)$.
It was, however, required to use a sufficiently large supercell
  in which the region of integration is well confined.
Furthermore, it was found that the integrals of the types
  $\int \sigma(\mathbf r) w_i(\mathbf r-\mathbf R_i) \sigma_\alpha(\mathbf r)d\mathbf r$ and
  $\int  \sigma^2(\mathbf r)w_i(\mathbf r-\mathbf R_i)d\mathbf r$
  show slow convergence with respect to the supercell size
  whereas the ratio
  $\int \sigma(\mathbf r) w_i(\mathbf r-\mathbf R_i) \sigma_\alpha(\mathbf r)d\mathbf r/\sqrt{\int  \sigma^2(\mathbf r)w_i(\mathbf r-\mathbf R_i)d\mathbf r}$
  converges rather quickly.
Hence sufficiently large supercells were used in the computation of $Z_{i\alpha}$ and $S_{i\alpha}$
  for the bulk solids, and it was confirmed that the computed values of $Z_{i\alpha}$ and $S_{i\alpha}$
  are independent of the size of the employed supercells.

\section{Results and Discussion}

In this section we first investigate the correlations between the energy difference $\mu$ and local (atom-partitioned) Carb\'{o} index.
We then explore the aforementioned interconnection between energetics and quantum similarity indices.
Next we employ the interpolation procedure developed in the preceding section 
  in order to devise a means for characterizing energetic heterogeneity of nanoparticles.
Finally we expound the similarity-based approach developed here
   by comparing the similarity-based energies to DFT-calculated energies
   for a number of unary C, Si, Pd, or Pt nanocrystals,
     uniformly mixed,
     core-shell segregated, and
     phase separated Pt-Pd nanoalloys, and
     Pt-C nanocompounds, cf. the {\it test\it} set.

\begin{figure}
  \centering
  \includegraphics[width=0.87\columnwidth,keepaspectratio=true]{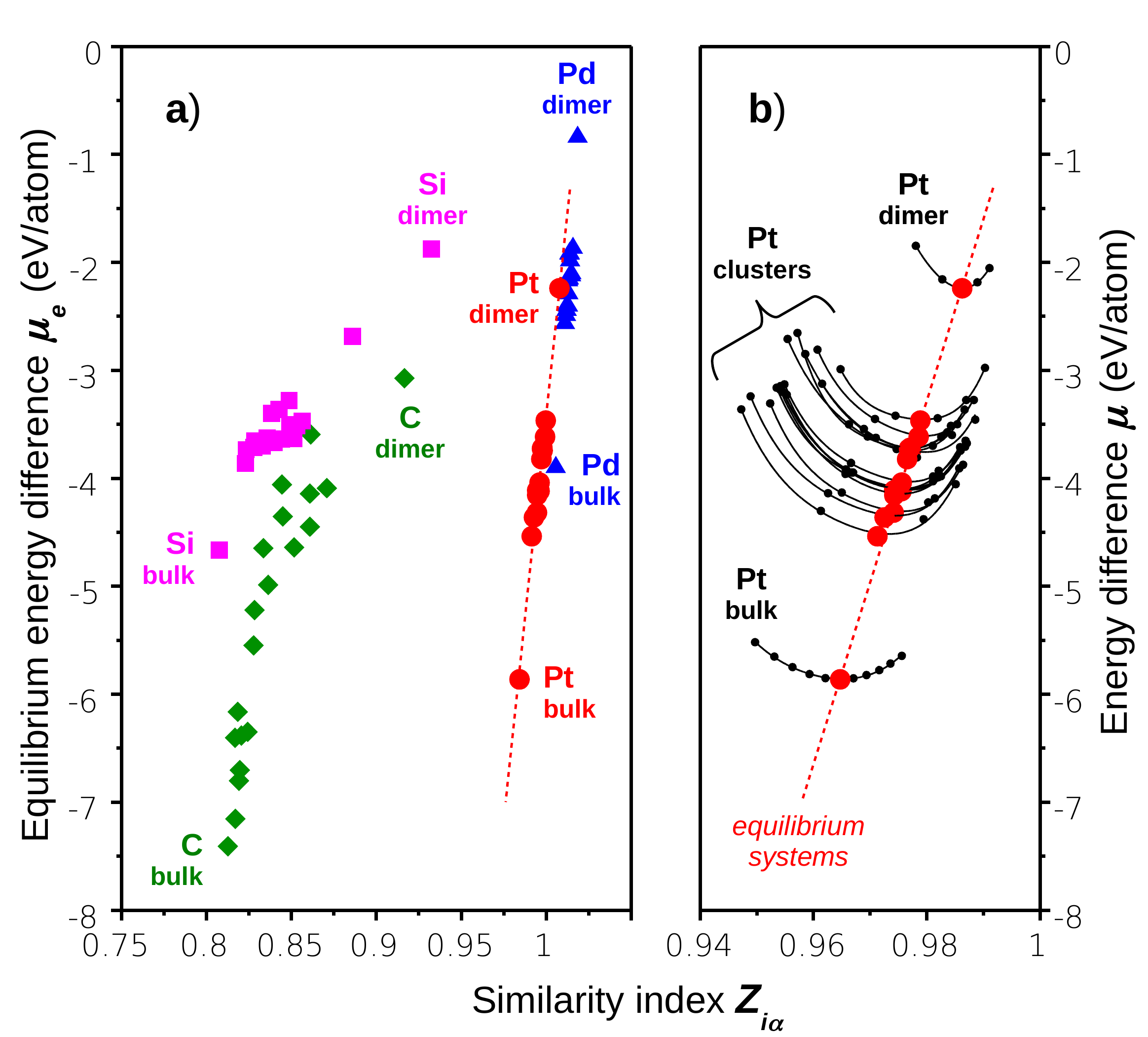}
  \caption{
  The {\it equilibrium\it} energy difference $\mu_{\rm e}$ (a)
  and energy difference $\mu$ (b)
  versus the atom-partitioned Carb\'{o} index $Z_{i\alpha}$
  for the unary {\it database\it} systems.
  The systems made of C, Si, Pd, and Pt atoms are represented
  by diamonds, squares, triangles, and circles, respectively.
  The points (a) or curves (b) corresponding to dimers and bulk solids
  are labeled while the unlabeled symbols represent the atomic clusters in the shape of regular polyhedra.
  }
  \label{MuvsZ}
\end{figure}

\subsection{Correlations between energy differences and similarity indices}
The plot of the {\it equilibrium\it} energy difference $\mu_e$
  versus the local Carb\'{o} index $Z_{i\alpha}$ is displayed in figure~1(a).
It is noticeable that there exists a correlation between $\mu_e$ and $Z_{i\alpha}$,
  which appears to be a distinct relationship for each atomic species.
Note that the correlation is seemingly {\it linear\it} for the Pd and Pt systems
  (as marked by the dashed line passing through the set of Pt systems).
Furthermore, it is seen that the set of Pd and Pt systems are grouped, i.e., they fall nearly on the same line.
For the Si systems the correlation could also be regarded {\it approximately\it} linear
  whereas the points for the C systems fall on a monotonic curve that is {\it not\it} linear and show a more pronounced scatter.
Yet, overall, there seems to exist a roughly one-to-one correspondence, i.e., bijection,
  between the {\it equilibrium\it} energy difference $\mu_e$ and $Z_{i\alpha}$,
  regardless of the atomic species.
This implies that the variety of local environments sampled by the {\it database\it} systems
  are adequately reflected by the atom-partitioned Carb\'{o} index.
On one hand,
this finding
  implying that a single number, viz. $Z_{i\alpha}$, per atom (as opposed to a function of space)  suffices to capture {\it energetics trends\it}
  is striking in the view of the holographic electron density theorem\cite{Mezey1999}
  which applies to the local electronic ground state density,
  i.e., a function of space restricted to some region (as opposed to a number).
It could, on the other hand, be expounded by noting that {\it similar\it} atoms
  (viz. atoms with close $Z_{i\alpha}$ values)
  would exhibit {\it similar\it} energetic stability
  (as indicated by close $\mu$ values).

\begin{figure}
  \centering
  \includegraphics[width=0.98\columnwidth,keepaspectratio=true]{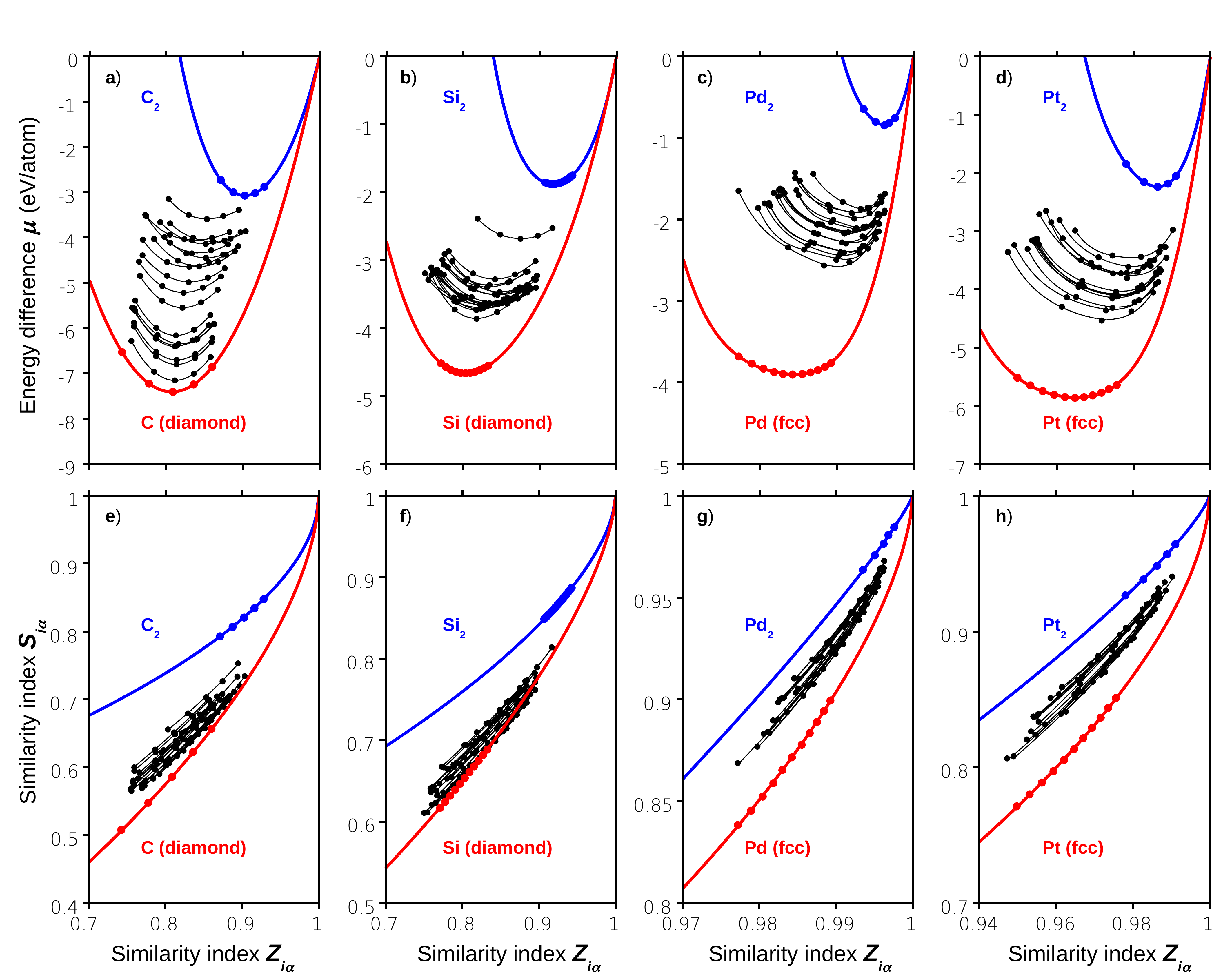}
  \caption{
  The energy difference $\mu$ (a)-(d)
  and the similarity index $S_{i\alpha}$ (e)-(h)
  versus the similarity index $Z_{i\alpha}$
  for the C (a) and (e), Si (b) and (f), Pd (c) and (g), and Pt (d) and (h)
  systems.
  The circles represent the calculation results to which the solid-line curves are fitted.
  The curves corresponding to dimers and bulk solids
  are labeled while the unlabeled symbols represent the atomic clusters in the shape of regular polyhedra.
  }
  \label{MUveSvsZ}
\end{figure}

\subsection{Energetics-similarity interconnection}
It is obviously interesting to see if the preceding analysis for the unstrained systems
  could as well be applied to the (negatively or positively) strained systems,
  i.e., if one could introduce a generalized $\mu$-$Z$ relationship.
Thus the plot of the energy difference $\mu$
  versus the atom-partitioned Carb\'{o} index $Z_{i\alpha}$ is drawn in figure~1(b)
  for the Pt systems,
  where the small (large) circles represents strained (equilibrium) systems.
Note that the large solid (red) circles as well as the (red) dashed lines on both panels of figure~1 are identical.
The small (black) circles represent the DFT-calculated ($\mu$,$Z_{i\alpha}$) values.
It is seen that the ($\mu$,$Z_{i\alpha}$) points fall on a distinct convex curve (represented by solid lines) for each system.
This observation is of practical significance,
  which makes it possible to parameterize $\mu$ as a function of $Z_{i\alpha}$.
In practice, this parameterization was carried out by
  using a distinct polynomial function 
  $\mu_i(Z_{i\alpha}) = C_0 + C_1 Z_{i\alpha} + C_2 Z^2_{i\alpha} + C_3 Z^3_{i\alpha} + C_4 Z^4_{i\alpha}$
  with a unique set of polynomial coefficients \{$C_k; k=0,1,2,3,4$\} for each system $i$,
  yielding the solid-line curves given in figure~1(b),
  which are obtained via fitting to the DFT-calculated points.
Repeating the same procedure for the C, Si, and Pd systems
  results in the $\mu$-$Z$ curves given in the top panels of figure~2
  where, for each atomic species, the curves for the atomic clusters lay necessarily between the curves for the dimer and bulk solid.
On the other hand, despite the utility of the $\mu$-$Z$ parameterization,
  there exists now {\it no\it} one-to-one correspondence
  between the energy difference $\mu$ and the atom-partitioned Carb\'{o} index,
  i.e., a given value of $Z_{i\alpha}$ does not correspond to a {\it unique\it} {\it database\it} system.
Thus one can utilize the atom-partitioned Carb\'{o} index as a measure of similarity {\it only\it} for equilibrium systems.
This, however, require {\it a priori\it} knowledge of the interatomic distances \{$d_e$\};
  in other words, the equilibrium geometries.
As mentioned above, this limitation is lifted by
devising
  a {\it second\it} indicator $S_{i\alpha}$ of local similarity
  given in equation~(\ref{eq2nd}).
Note that
  one would in principle need to use the local electronic ground state density\cite{Mezey1999}
  rather than numeric value of an integral of it, viz. $Z_{i\alpha}$,
  in order to differentiate closely-related systems,
  cf. the holographic electron density theorem.
Our analysis reveals that a pair of local similarity indices $(Z_{i\alpha},S_{i\alpha})$ 
  constitute an adequate chemical representation adopted here
  since a single similarity index fails
  to reflect {\it energetic trends\it}
  once the variations with the interatomic distance $d$ are taken into account.
The bottom panels of figure~2 show the plot of $S_{i\alpha}$
  versus $Z_{i\alpha}$ for the {\it database\it} systems.
Although the two similarity indices are not genuinely independent of each other, cf. equations~(\ref{eq1st}) and (\ref{eq2nd}),
  the $S_{i\alpha}$-$Z_{i\alpha}$ curves appear to be distinct for each system.
Furthermore
  all curves lie in the same region bounded by the dimer curve acting as an upper bound
  and the curve for the bulk solid, which serves as a lower bound.
Interestingly, one could make the same observation in the top panels of figure~2,
  where all the $\mu$-$Z$ curves are also bounded by the dimer and bulk curves.
Thus the $S$-$Z$ curves and the $\mu$-$Z$ curves are roughly ordered in a similar fashion.
This finding encourages one to utilize the {\it second\it} similarity index
  (in combination with the local Carb\'{o} index)
  to establish a bijection between the energy differences
  and similarity indices.
The latter is achieved by employing the {\it ad hoc\it} interpolation formula given in equation~(\ref{wii}),
  which enables one to obtain
  the contribution $\delta E_I$ of atom $I$ to the similarity-based total energy difference $\Delta E_{\rm sim}$
  from the similarity indices $(Z_{I\alpha},S_{I\alpha})$.

\begin{figure}
  \centering
  \includegraphics[width=0.98\columnwidth,keepaspectratio=true]{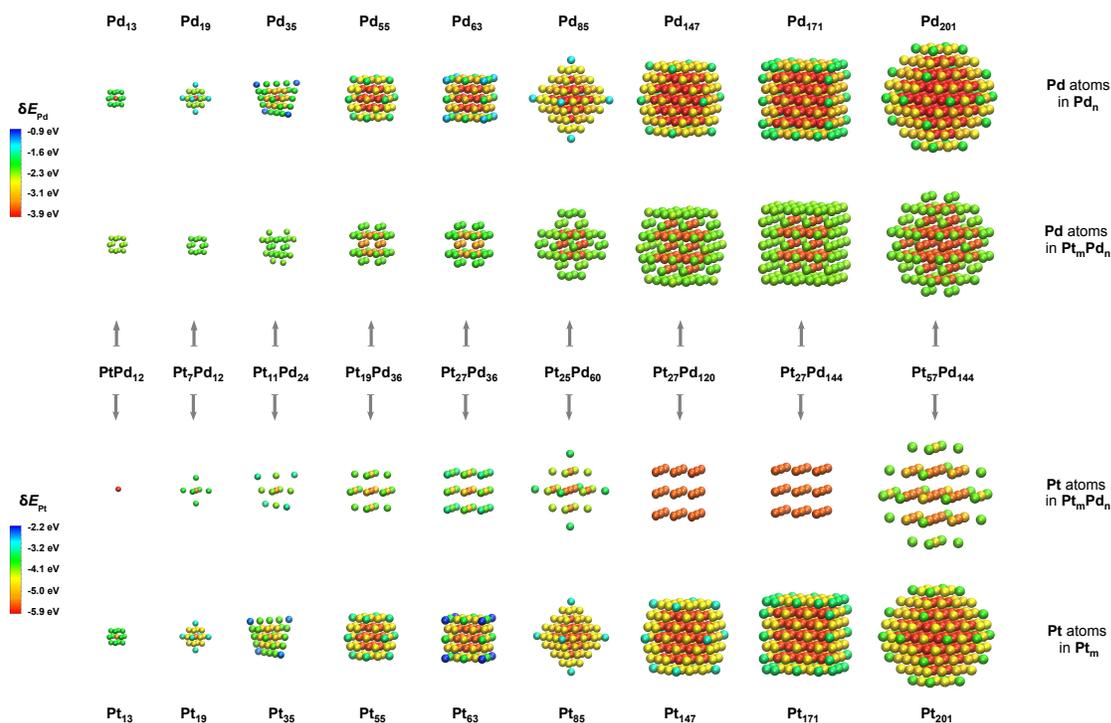}
  \caption{
  Color-coded graphs of $\delta E_I$ computed via equation~(\ref{eqEloc})
  for a series of Pd nanocrystals (top    graphs),
                  Pt-Pd nanoalloys   (middle graphs)
              and Pt    nanocrystals (bottom graphs).
  The scales on the left hand side represent the computed values
    of $\delta E_I$ for Pd atoms (top scale) and Pt atoms (bottom scale).
  }
  \label{renkli}
\end{figure}

\subsection{Means for characterizing energetic inhomogeneity}
Figure~3 displays color-coded graphs of $\delta E_I$ values
  for a series of Pd    nanocrystals (top    graphs),
                  Pt-Pd nanoalloys   (middle graphs)
              and Pt    nanocrystals (bottom graphs).
The scales on the left hand side represent the range for the computed values
  of $\delta E_I$ for Pd atoms (top scale) and Pt atoms (bottom scale).
The higher end (blue) of these scales corresponds to less stable (more reactive) atoms
  while the lower end (red) corresponds to more stable (less reactive) atoms.
It is encouraging to see that 
  the bulk-like atoms near to the center of a nanocrystal
  have $\delta E_{\rm Pd}$ and $\delta E_{\rm Pt}$ values around the lower end
  whereas $\delta E_{\rm Pd}$ and $\delta E_{\rm Pt}$ values
  are considerably higher for the low-coordinated atoms
  located on the faces, along the edges, or at the corners.
Thus using the $\delta E_I$ values
  facilitates the characterization of the energetic (site-specific, morphology-dependent) inhomogeneity
  of the nanocrystals.
Moreover {\it energetics trends\it} in regard to the size-dependence
  appears to be reasonable as one approach bulk-like energies in going from small to large nanocrystals.
Besides using the set of $\delta E_\alpha$ or $\delta E_\beta$ values
  enables one to look into local (e.g., site-specific) mixing of $\alpha$- and $\beta$-type atoms in a binary alloy formation.
For example, comparative inspection of Pt$_{35}$, Pt$_{11}$Pd$_{24}$, and Pd$_{35}$
  in figure~3
  shows that the Pt atoms at the corners of Pt$_{11}$Pd$_{24}$ are less energetic in comparison those of Pt$_{35}$,
  indicating that alloying Pt nanocrystal with Pd increases the energetic stability of the corner (Pt) atoms.
Since a similar analysis could be applied to any nanocrystal in a site-specific manner,
  using the set of $\delta E_I$ values would clearly be useful in elucidating trends
  in the size-, shape- and composition-dependent nanocrystal energetics.

\begin{figure}
  \centering
  \includegraphics[width=0.54\columnwidth,keepaspectratio=true]{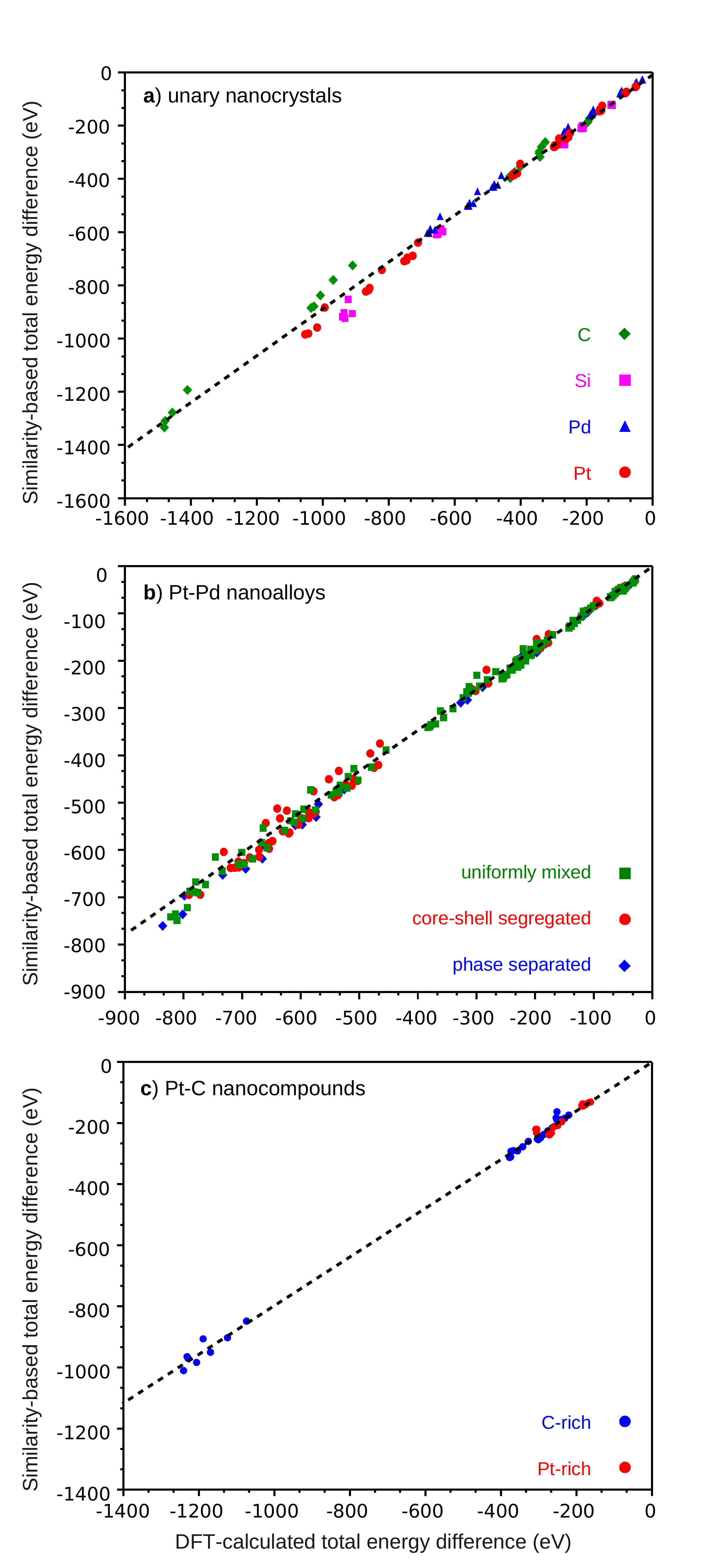}
  \caption{
  The similarity-based total energy difference $\Delta E_{\rm sim}$ introduced in equation~(\ref{eqEsim}) versus
  the   DFT-calculated total energy difference $\Delta E_{\rm DFT}$ given in equation~(\ref{eqEdft})
  for 
      a) the unary C, Si, Pd, or Pt nanocrystals,
      b) the Pt-Pd nanoalloys with various mixing patterns, and
      c) the Pt-C nanocompounds.
  }
  \label{EsimvsEdft}
\end{figure}

\subsection{Similarity-based energies}
Performing a sum over atoms as in equation~(\ref{eqEsim})
  enables one to obtain $\Delta E_{\rm sim}$ from the set of the atomic contributions $\delta E_I$,
  cf. figure~3.
It is then crucial to inquire if
  $\Delta E_{\rm sim}$
  could be utilized in lieu of $\Delta E_{\rm DFT}$ (calculated directly)
  for practical purposes, e.g., in the computer-aided design of nanocrystals.
Thus the plot of $\Delta E_{\rm sim}$ versus $\Delta E_{\rm DFT}$
  is drawn for the {\it test\it} systems in figure~4
  where the calculation results
  are included
  for a variety of nanocrystal sizes and shapes
  as well as a range of alloy/compound compositions with various mixing patterns.
Accordingly
  the upper, middle, and lower panels of figure~4 are devoted to
  the unary (C, Si, Pd, or Pt) nanocrystals,
  the Pt-Pd nanoalloys,
  the Pt-C nanocompounds, respectively.
Not only unstrained but also negatively or positively strained systems
  are included in these panels.
Here strained systems are characterized
  by the value of the interatomic distance $d$
  which is varied in the range of
  [1.37,1.55] \AA~ for the C  nanocrystals,
  [2.20,2.45] \AA~ for the Si nanocrystals,
  [2.57,3.11] \AA~ for the Pd nanocrystals,
  [2.57,2.91] \AA~ for the Pt nanocrystals,
  [2.55,3.12] \AA~ for the Pt-Pd nanoalloys,
  [1.90,2.25] \AA~ for the Pt-C nanocompounds.
A power-law regression analysis on the points marked by the filled symbols,
  which ensures that $\Delta E_{\rm sim} \rightarrow 0$ as $\Delta E_{\rm DFT} \rightarrow 0$,
  results in the dashed lines shown in the panels of figure~4,
  which is given by
  \begin{equation}
    \Delta E_{\rm sim} = - C \left [ -\Delta E_{\rm DFT} \right ]^\gamma \pm \delta,
    \label{eqFit}
  \end{equation}
  where both $\Delta E_{\rm sim}$ and $\Delta E_{\rm DFT}$ are in eV,
  and the coefficients $C$, the exponents $\gamma$, and the error bars $\delta$ are listed in table~1.
Note that the closeness of the values of $C$ and $\gamma$ to unity (in association with a small $\delta$)
  indicates that $\Delta E_{\rm sim}$ would follow the same {\it energetics trends\it} as $\Delta E_{\rm DFT}$.
On the other hand, having either $C$ or $\gamma$ smaller than unity
  is an indication that the similarity-based interpolation procedure
  results in {\it underestimation\it}.
The latter turns out to be the case
  as revealed by inspection of the slopes of the dashed lines in figure~4
  as well as the entries of table~1.
Yet the standard deviation $\delta$ is relatively small, i.e. on the order of 5\% in all cases,
  i.e., not only for the unary nanocrystals but also for the Pt-Pd nanoalloys and Pt-C nanocompounds.
In practice, one could invert equation~(\ref{eqFit}) in order to obtain the total energy differences
  more accurately for these systems.
It should be remarked that this description is not restricted to {\it equilibrium\it} configurations
 since a relatively wide range of interatomic distances are considered above.
Hence, portions of the potential energy surfaces of the nanocrystals are rendered accessible.
Further analysis reveals that
  the scatter of the points about the regression line
  is much less pronounced for slightly strained systems
  whereas significantly larger for the highly compressed systems.
One should consequently recognize that the similarity-based approach exemplified here
  would be more suitable in describing a portion of the potential energy surface
  that is in the vicinity of {\it equilibrium\it}, availability of which
  is clearly of great service in a multitude of design problems.


\begin{table}
  \centering
  \caption{\label{tbl1}
         The values for the coefficient $C$,
                        the exponent $\gamma$, and
                        the error bar $\delta$,
         which are introduced in equation~(\ref{eqFit}),
         for the {\it test\it} systems.
        }
\begin{tabular}{lccc}
\hline
                     & $C$       & $\gamma$  & $\delta$ (\%)  \\ \hline
  Unary nanocrystals & 1.015     & 0.981     & 5.9          \\ 
  Pt-Pd nanoalloys   & 0.929     & 0.989     & 5.9          \\ 
  Pt-C nanocompounds & 0.775     & 1.005     & 4.3          \\ 
\hline
\end{tabular}
\end{table}

It should be emphasized that
  $\Delta E_{\rm sim}$ for a {\it test\it} system is {\it truly\it} obtained
  by use of the (quantum) similarity of its atoms to the atoms of the {\it database\it} systems.
Recall that no member of the {\it test set\it} has been included
  in the fitting of the tabulated $\mu=\mu_i(Z_{i\alpha})$ relationships
  and the {\it database\it} systems are all {\it unary\it} systems.
It is thus remarkable that the similarity-based description of the energetics of binary nanoalloys is almost as good as 
  that of unary nanocrystals as seen by comparing figure~4(b) to figure~4(a).
The energetics of Pt-C nanocompounds constitute a greater challenge
  as evidenced by the relatively large deviation of the coefficient $C$ from unity,
  yet the correlation between $\Delta E_{\rm sim}$ and $\Delta E_{\rm DFT}$ for these systems
  is also significant, cf. figure~4(c), with a standard deviation $\delta < 5 \%$.
The latter gives evidence
  for the ubiquitous utility of the similarity-based approach developed here
  since the bonding characteristics of platinum carbide
  is peculiar\cite{Ivanovskii2009} owing to the mixed covalent-ionic-metallic interatomic interactions.

\section{Conclusion}

The results of the present investigation show that
  there exists an interconnection between
  the energy differences utilized in the modelling and simulation of nanocrystals and the quantum similarity measures,
  which is established here by devising two local similarity indices given in equations~(\ref{eq1st}) and (\ref{eq2nd}).
We show that this finding leads to the development of a new approach
  for computing the energy differences and total energies,
  which is based on the similarity of the nanocrystal atoms to the atoms of a set of {\it database\it} systems.
We find that the similarity-based energy differences
  exhibit the same trends as those obtained directly from the DFT calculations.
Subsequently, 
  it is demonstrated that
  our similarity-based approach could be used to explore the size-, shape- and composition-dependent nanocrystal energetics.
It should be remarked that {\it no\it} knowledge of the {\it equilibrium\it} geometries is needed {\it a priori\it} in this approach
  since it is sufficiently general to cover non-equilibrium configurations.
In particular, it is intriguing that
  the similarity-based description of energetics of the binary systems
  is nearly as good as that of unary systems
  -albeit {\it no\it} binary systems are included among the {\it database\it} systems.
Furthermore, it is demonstrated that
  our similarity-based approach
  provides a means for characterizing the energetic inhomogeneity of the nanocrystals, cf. figure~3.
Accordingly, 
  we project that the methodology of this paper
  would be of great service in informatics-driven approaches, 
  i.e. materials informatics\cite{rajan05}.
Besides, we  anticipate that the similarity-based approach presented here could be generalized to explore other physicochemical quantities,
  e.g. (site-specific) adsorption energies, and could then be utilized in the surface engineering of nanocrystals.

\ack

This work was supported by TUBITAK under Grant No. TBAG-109T677.
The computations were carried out at the High Performance and Grid Computing Center (TRUBA resources) of TUBITAK ULAKBIM.


\bibliographystyle{unsrt}

\begin{thebibliography}{}

\end{thebibliography}


\section*{References}

\begin{thebibliography}{40}

\bibitem{Trindade2001}
Trindade T, O'Brien P and Pickett NL {2001}
\newblock {Nanocrystalline semiconductors: Synthesis, properties and perspectives}
\newblock {\em {Chem. Mat.}} {13} {3843--3858}

\bibitem{Ferrando2008}
Ferrando R, Jellinek J and Johnston RL {2008}
\newblock {Nanoalloys: From theory to applications of alloy clusters and nanoparticles}
\newblock {\em {Chem. Rev.}} {108} {845--910}

\bibitem{Xia2009}
Xia Y, Xiong Y, Lim B and Skrabalak SE {2009}
\newblock {Shape-controlled synthesis of metal nanocrystals: Simple chemistry meets complex physics?}
\newblock {\em {Angew. Chem.-Int. Edit.}} {48} {60--103}

\bibitem{Willett1998}
Willett P, Barnard JM and Downs GM {1998}
\newblock {Chemical similarity searching}
\newblock {\em {J. Chem. Inf. Comput. Sci.}} {38} {983--996}

\bibitem{Sheridan2002}
Sheridan RP and Kearsley SK {2002}
\newblock {Why do we need so many chemical similarity search methods?}
\newblock {\em {Drug Discov. Today}} {7} {903--911}

\bibitem{Engel2006}
Engel T {2006}
\newblock {Basic overview of chemoinformatics}
\newblock {\em {J. Chem. Inf. Model.}} {46} {2267--2277}

\bibitem{Carbo1980}
Carb\'{o} R, Leyda L and Arnau M {1980}
\newblock {How similar is a molecule to another - an electron-density measure of similarity between 2 molecular-structures}
\newblock {\em {Int. J. Quant. Chem.}} {17} {1185--1189}

\bibitem{Mezey1999}
Mezey PG {1999}
\newblock {The holographic electron density theorem and quantum similarity measures}
\newblock {\em {Mol. Phys.}} {96} {169--178}

\bibitem{Chermette1999}
Chermette H {1999}
\newblock {Chemical reactivity indexes in density functional theory}
\newblock {\em {J. Comput. Chem.}} {20} {129--154}

\bibitem{Karelson1996}
Karelson M, Lobanov VS and Katritzky AR {1996}
\newblock {Quantum-chemical descriptors in QSAR/QSPR studies}
\newblock {\em Chem. Rev.} {96} {1027--1043}

\bibitem{Bultinck2005}
Bultinck P and Carb\'{o}-Dorca R {2005}
\newblock {Molecular quantum similarity using conceptual DFT descriptors}
\newblock {\em {J. Chem. Sci.}} {117} {425--435}

\bibitem{Geerlings2005}
Geerlings P, Boon G, Van Alsenoy C and De Proft F {2005}
\newblock {Density functional theory and quantum similarity}
\newblock {\em Int. J. Quant. Chem.} {101} {722--732}

\bibitem{Boon2001}
Boon G, Langenaeker W, De Proft F, De Winter H, Tollenaere JP and Geerlings P {2001}
\newblock {Systematic study of the quality of various quantum similarity descriptors. Use of the autocorrelation function and principal component analysis}
\newblock {\em {J. Phys. Chem. A}} {105} {8805--8814}

\bibitem{Good1993}
Good AC, Peterson SJ and Richards WG {1993}
\newblock {QSARs from similarity-matrices - technique validation and application in the comparison of different similarity evaluation methods}
\newblock {\em {J. Med. Chem.}} {36} {2929--2937}

\bibitem{Besalu2002}
Besal\'{u} E, Girones X, Amat L and Carb\'{o}-Dorca R {2002}
\newblock {Molecular quantum similarity and the fundamentals of QSAR}
\newblock {\em {Accounts Chem. Res.}} {35} {289--295}

\bibitem{Amat2001}
Amat L, Besal\'{u} E, Carb\'{o}-Dorca R and Ponec R {2001}
\newblock {Identification of active molecular sites using quantum-self-similarity measures}.
\newblock {\em {J. Chem. Inform. Comput. Sci.}} {41} {978--991}

\bibitem{Rouvray1992}
Rouvray DH {1992}
\newblock {Definition and role of similarity concepts in the chemical and physical sciences}
\newblock {\em J. Chem. Inf. Comput. Sci.} {32} {580--586}

\bibitem{Bartok2013}
Bart\'ok AP, Kondor R and Cs\'anyi G {2013}
\newblock {On representing chemical environments}
\newblock {\em Phys. Rev. B} {87} {184115}

\bibitem{Hirshfeld1977}
Hirshfeld FL {1977}
\newblock {Bonded-atom fragments for describing molecular charge-densities}
\newblock {\em {Theor. Chem. Acc.}} {44} {129--138}

\bibitem{DeProft2003}
De Proft F, Vivas-Reyes R, Peeters A, Van Alsenoy C and Geerlings P {2003}
\newblock {Hirshfeld partitioning of the electron density: Atomic dipoles and their relation with functional group properties}
\newblock {\em {J. Comput. Chem.}} {24} {463--469}

\bibitem{Parr1983}
Parr RG and  Bartolotti LJ {1983}
\newblock {Some remarks on the density functional theory of few-electron systems}
\newblock {\em {J. Phys. Chem.}} {87} {2810--2815}

\bibitem{Ayers2000}
Ayers PW {2000}
\newblock {Density per particle as a descriptor of Coulombic systems}
\newblock {\em {Proc. Natl. Acad. Sci. USA}} {97} {1959--1964}

\bibitem{Proft2004}
De Proft F, Ayers PW, Sen KD and Geerlings P {2004}
\newblock {On the importance of the ``density per particle'' (shape function) in the density functional theory}
\newblock {\em {J. Chem. Phys.}} {120} {9969-9973}

\bibitem{Kilic2010}
K{\i}l{\i}\c{c} \c{C} {2010}
\newblock {Size- and shape-dependent energetics of transition-metal nanocrystals}
\newblock {\em {Solid State Commun.}} {150} {2333--2336}

\bibitem{Kilic2011}
K{\i}l{\i}\c{c} \c{C} {2011}
\newblock {Energy-distance relation for fcc transition metal nanocrystals}
\newblock {\em EPL} {93} {26004}

\bibitem{Behler2011}
Behler J {2011}
\newblock {Neural network potential-energy surfaces in chemistry: a tool for large-scale simulations}
\newblock {\em Phys. Chem. Chem. Phys.} 13 17930--17955

\bibitem{CRC}
Lide D {2008}
\newblock {\em CRC handbook of chemistry and physics : a ready-reference book of chemical and physical data}
\newblock {(Boca Raton, FL: CRC)}

\bibitem{Parretti1997}
Parretti MF, Kroemer RT, Rothman JH and Richards WG {1997}
\newblock {Alignment of molecules by the monte carlo optimization of molecular similarity indices}
\newblock {\em J. Comput. Chem.} {18} {1344--1353}

\bibitem{Bultinck2003}
Bultinck P, Kuppens T, Giron\'{e}s X and Carb\'{o}-Dorca R {2003}
\newblock {Quantum similarity superposition algorithm (QSSA): A consistent scheme for molecular alignment and molecular similarity based on quantum chemistry}
\newblock {\em J. Chem. Inf. Comput. Sci.} 43 1143--1150

\bibitem{Bultinck03}
Bultinck P, Carb\'{o}-Dorca R and Van Alsenoy C {2003}
\newblock {Quality of approximate electron densities and internal consistency of molecular alignment algorithms in molecular quantum similarity}
\newblock {\em J. Chem. Inform. Comput. Sci.} {43} {1208--1217}

\bibitem{Ono2005}
Ono S, Kikegawa T and Ohishi Y {2005}
\newblock {A high-pressure and high-temperature synthesis of platinum carbide}
\newblock {\em {Solid State Commun.}} {133} {55--59}

\bibitem{Ivanovskii2009}
Ivanovskii AL {2009}
\newblock {Platinum group metal nitrides and carbides: synthesis, properties and simulation}
\newblock {\em {Russ. Chem. Rev.}} {78} {303--318}

\bibitem{PBE}
Perdew JP, Burke K and Ernzerhof M {1996}
\newblock {Generalized gradient approximation made simple}
\newblock {\em Phys. Rev. Lett.} {77} {3865--3868}

\bibitem{PAW}
Blochl PE {1994}
\newblock {Projector augmented-wave method}
\newblock {\em {Phys. Rev. B}} {50} {17953--17979}

\bibitem{VASP}
Kresse G and F\"{u}rthmuller J {1996}
\newblock {Efficiency of ab-initio total energy calculations for metals and semiconductors using a plane-wave basis set}
\newblock {\em {Comput. Mater. Sci.}} {6} {15--50}

\bibitem{VASPpaw}
Kresse G and Joubert D {1999}
\newblock {From ultrasoft pseudopotentials to the projector augmented-wave method}
\newblock {\em {Phys. Rev. B}} {59} {1758--1775}

\bibitem{VASPchem}
Sun GY, Kurti J, Rajczy P, Kertesz M, Hafner J and Kresse G {2003}
\newblock {Performance of the Vienna ab initio simulation package (VASP) in chemical applications}
\newblock {\em {J. Mol. Struct. - THEOCHEM}} {624} {37--45}

\bibitem{MP1976}
Monkhorst HJ and Pack JD {1976}
\newblock {Special points for brillouin-zone integrations}
\newblock {\em Phys. Rev. B} {13} {5188--5192}

\bibitem{dcuhre}
Berntsen J, Espelid TO and Genz A {1991}
\newblock {Algorithm 698: Dcuhre: an adaptive multidemensional integration routine for a vector of integrals}
\newblock {\em ACM Trans. Math. Softw.} {17} {452--456}

\bibitem{spline}
Boor C {2001}
\newblock {\em A practical guide to splines}
\newblock {(New York: Springer)}

\bibitem{rajan05}
Rajan K {2005}
\newblock {Materials informatics}
\newblock {\em Mater. Today} {8} {38--45}

\end{thebibliography}

\clearpage

\section{Supplementary Data}

\setcounter{table}{0}

\begin{table}[h]
\caption{ 
         The coefficients in the polynomial relationship
         $\mu_i(Z_{i\alpha})=C_0+C_1 Z_{i\alpha}+ C_2 Z^2_{i\alpha} + C_3 Z^3_{i\alpha} + C_4 Z^4_{i\alpha}$
         for the {\it database\it} systems with $\alpha=$ C.
        }
\begin{tabular}{lrrrrr}\hline
\multicolumn{1}{l}{$i$} & \multicolumn{1}{c}{$C_0$} & \multicolumn{1}{c}{$C_1$} & \multicolumn{1}{c}{$C_2$} & \multicolumn{1}{c}{$C_3$} & \multicolumn{1}{c}{$C_4$} \\ \hline
                               Dimer  &    -3.073 & -0.020& 317.605 & -662.128 & 7537.589  \\
                         Tetrahedron  &    -3.585 & -0.243 & 131.127 & -681.23 & 6174.054  \\ 
                          Octahedron  &    -4.448 & -0.095 & 142.577 & -87.009 & 3321.910 \\ 
                                Cube  &    -5.212 & -0.223 & 166.607 & -491.447 & 2717.141  \\ 
                         Icosahedron  &    -4.089 & -0.137 & 117.769 & 60.221 & 4425.848  \\ 
                        Dodecahedron  &    -6.384 & -0.008 & 221.448 & -283.245 & 491.922  \\ 
                       Cuboctahedron  &    -4.335 & -1.519 & 144.830& -354.519 & 3219.137  \\ 
                   Icosidodecahedron  &    -4.978 & -0.373 & 145.124 & -97.645 & 1514.808  \\ 
              Rhombicosidodecahedron  &    -4.641 & -0.109 & 125.075 & -684.302 & 4934.909  \\ 
                 Rhombicuboctahedron  &    -4.639 & -0.068 & 159.520& -129.355 & 2021.972  \\ 
                           Snub cube  &    -4.140 & -0.142 & 122.413 & -7.260 & 3077.307  \\ 
                   Snub dodecahedron  &    -4.056 & -0.205 & 95.873 & -328.335 & 4150.783  \\ 
                      Truncated cube  &    -6.163 & -0.008 & 245.263 & -493.121 & 651.689  \\ 
             Truncated cuboctahedron  &    -6.704 & -0.003 & 242.436 & -347.009 & 452.461  \\ 
              Truncated dodecahedron  &    -6.403 & -0.003 & 262.507 & -577.442 & 735.888  \\ 
               Truncated icosahedron  &    -7.155 & -0.001 & 250.095 & -314.456 & 294.125  \\ 
         Truncated icosidodecahedron  &    -6.803 & -0.001 & 250.759 & -371.083 & 398.845  \\ 
                Truncated octahedron  &    -6.346 & -0.146 & 220.742 & -3.372 & -833.696  \\ 
               Truncated tetrahedron  &    -5.547 & -0.038 & 203.429 & -369.196 & 1140.742  \\ 
                Bulk solid (diamond)  &    -7.408 & -0.019 & 201.196 & -77.268 & 364.538  \\ 
\hline
\end{tabular}
\end{table}

\begin{table}
\caption{
         The coefficients in the polynomial relationship
         $\mu_i(Z_{i\alpha})=C_0+C_1 Z_{i\alpha}+ C_2 Z^2_{i\alpha} + C_3 Z^3_{i\alpha} + C_4 Z^4_{i\alpha}$
         for the {\it database\it} systems with $\alpha=$ Si.
        }
\begin{tabular}{lrrrrr}\hline
\multicolumn{1}{l}{$i$} & \multicolumn{1}{c}{$C_0$} & \multicolumn{1}{c}{$C_1$} & \multicolumn{1}{c}{$C_2$} & \multicolumn{1}{c}{$C_3$} & \multicolumn{1}{c}{$C_4$} \\ \hline
                               Dimer  &    -1.878 & 0.019 & 199.688 & -304.137 & 14483.305  \\
                         Tetrahedron  &    -2.682 & -0.164 & 79.253 & 66.414 & 5340.125  \\ 
                          Octahedron  &    -3.510& -0.127 & 84.063 & -21.272 & 2699.598  \\ 
                                Cube  &    -3.468 & -0.114 & 100.733 & -98.621 & 2757.457  \\ 
                         Icosahedron  &    -3.629 & -0.158 & 76.853 & 30.174 & 2700.339  \\ 
                        Dodecahedron  &    -3.673 & 0.260& 85.190& -350.766 & 3664.758  \\ 
                       Cuboctahedron  &    -3.501 & -0.135 & 84.383 & -53.555 & 2500.681  \\ 
                   Icosidodecahedron  &    -3.647 & -0.075 & 92.153 & -126.553 & 2048.645  \\ 
              Rhombicosidodecahedron  &    -3.622 & -0.183 & 85.326 & -77.365 & 1782.164  \\ 
                 Rhombicuboctahedron  &    -3.636 & -0.018 & 74.042 & -10.420& 2377.063  \\ 
                           Snub cube  &    -3.702 & -0.06 & 65.252 & -56.723 & 2221.128  \\ 
                   Snub dodecahedron  &    -3.650& -0.178 & 74.868 & -52.119 & 1598.565  \\ 
                      Truncated cube  &    -3.366 & 0.036 & 92.328 & -439.824 & 3800.469  \\ 
             Truncated cuboctahedron  &    -3.719 & 0.012 & 118.783 & -526.290& 2871.248  \\ 
              Truncated dodecahedron  &    -3.396 & -0.101 & 95.441 & -254.356 & 2349.424  \\ 
               Truncated icosahedron  &    -3.858 & -0.121 & 139.553 & -682.515 & 3016.681  \\ 
         Truncated icosidodecahedron  &    -3.732 & -0.133 & 122.766 & -747.689 & 3783.626  \\ 
                Truncated octahedron  &    -3.654 & -0.040& 111.444 & -98.950& 1228.149  \\ 
               Truncated tetrahedron  &    -3.277 & -0.104 & 95.328 & -198.807 & 2581.799  \\ 
                Bulk solid (diamond)  &    -4.664 & -0.005 & 132.739 & -349.581 & 1444.348  \\ 
\hline
\end{tabular}
\end{table}

\begin{table}
\caption{
         The coefficients in the polynomial relationship
         $\mu_i(Z_{i\alpha})=C_0+C_1 Z_{i\alpha}+ C_2 Z^2_{i\alpha} + C_3 Z^3_{i\alpha} + C_4 Z^4_{i\alpha}$
         for the {\it database\it} systems with $\alpha=$ Pd.
        }
\begin{tabular}{lrrrrr}\hline
\multicolumn{1}{l}{$i$} & \multicolumn{1}{c}{$C_0$} & \multicolumn{1}{c}{$C_1$} & \multicolumn{1}{c}{$C_2$} & \multicolumn{1}{c}{$C_3$} & \multicolumn{1}{c}{$C_4$} \\ \hline
                               Dimer  &    -0.840& -6.400 & 33311.393 & 3721206.184 & 511101393.400 \\ 
                         Tetrahedron  &    -1.873 & -2.569 & 16389.493 & 2770276.589 & 261971715.68  \\ 
                          Octahedron  &    -2.117 & -16.320& 6705.320& 2202612.455 & 273801033.220 \\ 
                                Cube  &    -2.099 & -16.802 & 9114.439 & 2498668.588 & 251443464.030  \\ 
                         Icosahedron  &    -2.391 & -16.539 & 5770.537 & 1412107.506 & 134315549.540  \\ 
                        Dodecahedron  &    -2.427 & -16.672 & 7362.723 & 2456123.506 & 341325123.540  \\ 
                       Cuboctahedron  &    -2.279 & -15.448 & 6237.267 & 1335147.165 & 149468839.650  \\ 
                   Icosidodecahedron  &    -2.268 & -13.721 & 6534.375 & 1066636.190& 116535424.090  \\ 
              Rhombicosidodecahedron  &    -2.436 & -12.002 & 6829.432 & 1021713.153 & 95175168.583  \\ 
                 Rhombicuboctahedron  &    -2.387 & -14.826 & 5718.681 & 1189598.403 & 121041265.850  \\ 
                           Snub cube  &    -2.475 & -17.289 & 4257.951 & 1043419.391 & 104446948.320  \\ 
                   Snub dodecahedron  &    -2.569 & -15.577 & 7444.008 & 890949.576 & 57481732.306  \\ 
                      Truncated cube  &    -1.907 & -13.208 & 5183.779 & 1300439.300& 201545862.790  \\ 
             Truncated cuboctahedron  &    -2.143 & -13.097 & 5912.355 & 1114780.637 & 142341680.100  \\ 
              Truncated dodecahedron  &    -1.906 & -11.746 & 8052.126 & 1226154.160& 163401620.030  \\ 
               Truncated icosahedron  &    -2.164 & -11.775 & 6970.046 & 1165244.069 & 123592942.310  \\ 
         Truncated icosidodecahedron  &    -2.148 & -11.657 & 6461.044 & 1068337.858 & 128956261.350  \\ 
                Truncated octahedron  &    -2.144 & -11.828 & 6638.110& 1336198.448 & 150585430.490  \\ 
               Truncated tetrahedron  &    -1.977 & -11.391 & 8351.823 & 1444832.497 & 202017269.710  \\ 
                    Bulk solid (fcc)  &    -3.901 & -3.186 & 4385.080& 265509.255 & 30014956.142  \\ 
\hline
\end{tabular}
\end{table}

\begin{table}
\caption{
         The coefficients in the polynomial relationship
         $\mu_i(Z_{i\alpha})=C_0+C_1 Z_{i\alpha}+ C_2 Z^2_{i\alpha} + C_3 Z^3_{i\alpha} + C_4 Z^4_{i\alpha}$
         for the {\it database\it} systems with $\alpha=$ Pt.
        }
\begin{tabular}{lrrrrr}\hline
\multicolumn{1}{l}{$i$} & \multicolumn{1}{c}{$C_0$} & \multicolumn{1}{c}{$C_1$} & \multicolumn{1}{c}{$C_2$} & \multicolumn{1}{c}{$C_3$} & \multicolumn{1}{c}{$C_4$} \\ \hline
                               Dimer  &    -2.239 & -1.024 & 6923.321 & 227192.717 & 9823552.539  \\ 
                          Octahedron  &    -3.163 & -1.457 & 2765.467 & 81231.426 & 9123526.461  \\
                          Octahedron  &    -3.455 & -1.794 & 1539.467 & 89757.477 & 9766553.492  \\ 
                                Cube  &    -3.745 & -1.180 & 2664.201 & 110609.121 & 9235782.643  \\ 
                         Icosahedron  &    -3.965 & -1.680 & 6354.282 &  98651.463 & 8230902.315  \\ 
                        Dodecahedron  &    -3.565 & -1.525 & 9584.201 &  92629.274 & 9256123.345  \\ 
                       Cuboctahedron  &    -3.739 & -10.083 & 1016.310 & 80505.616 & 7794825.247  \\ 
                   Icosidodecahedron  &    -4.087 & -5.506 & 1700.004 & 99336.991 & 4531031.341  \\ 
              Rhombicosidodecahedron  &    -4.301 & -6.027 & 1741.367 & 85256.841 & 4067693.288  \\ 
                 Rhombicuboctahedron  &    -4.105 & -5.443 & 2020.654 & 100834.379 & 4306097.006  \\ 
                           Snub cube  &    -4.338 & -7.474 & 1609.428 & 73361.770 & 3170892.929  \\ 
                   Snub dodecahedron  &    -4.509 & -7.448 & 1604.538 & 63718.292 & 2772333.570  \\ 
                      Truncated cube  &    -3.719 & -3.514 & 2780.603 & 93706.049 & 3989379.764  \\ 
             Truncated cuboctahedron  &    -4.095 & -5.643 & 2040.338 & 72807.266 & 3718101.004  \\ 
              Truncated dodecahedron  &    -3.717 & -0.827 & 2721.354 & 90423.545 & 4418129.817  \\ 
               Truncated icosahedron  &    -4.100 & -6.002 & 1858.331 & 81242.419 & 4454911.798  \\ 
         Truncated icosidodecahedron  &    -4.140 & -5.594 & 2066.221 & 74635.086 & 3730411.533  \\ 
                Truncated octahedron  &    -4.027 & -6.057 & 1886.612 & 98087.942 & 4539336.168  \\ 
               Truncated tetrahedron  &    -3.603 & -6.091 & 2304.512 & 127671.403 & 6669393.171  \\ 
                    Bulk solid (fcc)  &    -5.858 & -1.159 & 1463.681 & 28636.243 & 1814989.886  \\
\hline
\end{tabular}
\end{table}

\clearpage

\begin{table}
\caption{Unary (X=C or Si; Y=Pd or Pt) nanocrystals employed in this study.}
\begin{tabular}{ll}\hline
\textbf{Nanocrystal}  & \textbf{Shape}  \\
\hline
X$_{35}$      &Octahedron            \\
X$_{51}$      &Truncated tetrahedron \\
X$_{59}$      &Truncated tetrahedron \\
X$_{75}$      &Truncated cube        \\
X$_{165}$     &Octahedron        \\   
X$_{239}$     &Truncated cube        \\

Y$_{13}$      &Cuboctahedron      \\
Y$_{19}$      &Octahedron       \\
Y$_{35}$      &Tetrahedron       \\
Y$_{55}$      &Cuboctahedron       \\
Y$_{63}$      &Cube        \\
Y$_{85}$      &Octahedron      \\
Y$_{147}$     &Cuboctahedron      \\
Y$_{171}$     &Cube         \\
Y$_{201}$     &Truncated octahedron \\
\hline
\end{tabular}
\end{table}

\clearpage

\begin{table}
\caption{Uniformly mixed nanoalloys employed in this study.}
\begin{tabular}{lll}\hline
\textbf{} & \textbf{Pt Composition} & \textbf{Shape}  \\
\hline
Pt$_{12}$  Pd         &0.923  &Cuboctahedron        \\ 
Pt$_{12}$  Pd$_{7}$   &0.632  &Octahedron           \\ 
Pt$_{24}$  Pd$_{11}$  &0.686  &Tetrahedron          \\ 
Pt$_{36}$  Pd$_{19}$  &0.655  &Cuboctahedron        \\ 
Pt$_{36}$  Pd$_{27}$  &0.571  &Cube                 \\ 
Pt$_{60}$  Pd$_{25}$  &0.706  &Octahedron           \\ 
Pt$_{120}$ Pd$_{27}$  &0.816  &Cuboctahedron        \\ 
Pt$_{144}$ Pd$_{27}$  &0.842  &Cube                 \\ 

PtPd$_{12}$           &0.077  &Cuboctahedron        \\ 
Pt$_{7}$   Pd$_{12}$  &0.368  &Octahedron           \\   
Pt$_{11}$  Pd$_{24}$  &0.314  &Tetrahedron          \\ 
Pt$_{19}$  Pd$_{36}$  &0.345  &Cuboctahedron        \\ 
Pt$_{27}$  Pd$_{36}$  &0.429  &Cube                 \\ 
Pt$_{25}$  Pd$_{60}$  &0.294  &Octahedron           \\  
Pt$_{27}$  Pd$_{120}$ &0.184  &Cuboctahedron        \\ 
Pt$_{27}$  Pd$_{144}$ &0.158  &Cube                 \\ 
Pt$_{57}$  Pd$_{144}$ &0.284  &Truncated octahedron \\ 
\hline
\end{tabular}
\end{table}

\clearpage

\begin{table}
\caption{Core-shell segregated nanoalloys employed in this study.}
\begin{tabular}{lll}\hline
\textbf{} & \textbf{Pt Composition} & \textbf{Shape}  \\
\hline
PtPd$_{18}$           &0.053  &Octahedron           \\ 
PtPd$_{34}$           &0.029  &Tetrahedron          \\
Pt$_{13}$Pd$_{42}$    &0.236  &Cuboctahedron        \\
Pt$_{13}$Pd$_{50}$    &0.206  &Cube	                \\
Pt$_{19}$Pd$_{66}$    &0.224  &Octahedron           \\  
Pt$_{13}$Pd$_{134}$   &0.088  &Cuboctahedron        \\
Pt$_{19}$Pd$_{128}$   &0.129  &Cuboctahedron        \\
Pt$_{55}$Pd$_{92}$    &0.374  &Cuboctahedron        \\
Pt$_{158}$Pd$_{13}$   &0.924  &Cube                 \\
Pt$_{152}$Pd$_{19}$   &0.889  &Cube                 \\
Pt$_{136}$Pd$_{35}$   &0.795  &Cube                 \\
Pt$_{116}$Pd$_{55}$   &0.678  &Cube                 \\
Pt$_{108}$Pd$_{63}$   &0.632  &Cube                 \\
Pt$_{188}$Pd$_{13}$   &0.935  &Truncated octahedron \\
Pt$_{182}$Pd$_{19}$   &0.905  &Truncated octahedron \\
Pt$_{166}$Pd$_{35}$   &0.826  &Truncated octahedron \\
Pt$_{146}$Pd$_{55}$   &0.726  &Truncated octahedron \\
\hline
\end{tabular}
\end{table}

\clearpage

\begin{table}
\caption{Phase separated nanoalloys employed in this study.}
\begin{tabular}{lll}
\hline
\textbf{} & \textbf{Pt Composition} & \textbf{Shape}  \\
\hline
Pt$_{4} $Pd$_{9}$     &0.308   &Cuboctahedron	     \\
Pt$_{5} $Pd$_{14}$    &0.263   &Octahedron	     \\
Pt$_{13}$Pd$_{22}$    &0.371   &Tetrahedron	     \\
Pt$_{21}$Pd$_{34}$    &0.382   &Cuboctahedron	     \\
Pt$_{25}$Pd$_{38}$    &0.397   &Cube		     \\
Pt$_{30}$Pd$_{55}$    &0.353   &Octahedron	     \\
Pt$_{61}$Pd$_{86}$    &0.415   &Cuboctahedron	     \\
Pt$_{73}$Pd$_{98}$    &0.427   &Cube		     \\
Pt$_{119}$Pd$_{82}$   &0.592   &Truncated octahedron \\
\hline
\end{tabular}
\end{table}

\clearpage

\begin{table}
\caption{Zincblende nanocompounds employed in this study.}
\begin{tabular}{lll}
\hline
\textbf{} & \textbf{Pt Composition} & \textbf{Shape}  \\
\hline
Pt$_{16}$C$_{19}$     &0.457   &Octahedron            \\
Pt$_{20}$C$_{31}$     &0.392   &Truncated tetrahedron \\
Pt$_{28}$C$_{31}$     &0.475   &Truncated tetrahedron \\
Pt$_{32}$C$_{43}$     &0.427   &Truncated cube        \\
Pt$_{80}$C$_{85}$     &0.485   &Octahedron	      \\   
Pt$_{104}$C$_{135}$   &0.435   &Truncated cube        \\
Pt$_{19}$C$_{16}$     &0.543   &Octahedron	      \\
Pt$_{31}$C$_{20}$     &0.608   &Truncated tetrahedron \\
Pt$_{31}$C$_{28}$     &0.525   &Truncated tetrahedron \\
Pt$_{43}$C$_{32}$     &0.573   &Truncated cube        \\
Pt$_{85}$C$_{80}$     &0.515   &Octahedron	      \\
Pt$_{135}$C$_{104}$   &0.565   &Truncated cube        \\
\hline
\end{tabular}
\end{table}

\end{document}